\documentclass[aps,prb,singlecolumn,superscriptaddress]{revtex4}
\usepackage{amsfonts}
\usepackage{amssymb}
\usepackage{amsmath}
\usepackage{graphicx}
\usepackage{epsfig}
\usepackage{color}
\usepackage{amsmath,bm}

\begin{document}

\title{Insulating-to-conducting state transition in the bilayer Hubbard
model induced by a perpendicular quench field}
\author{X. Z. Zhang}
\affiliation{College of Physics and Materials Science, Tianjin Normal University, Tianjin
300387, China}
\author{Z. Song}
\email{songtc@nankai.edu.cn}
\affiliation{School of Physics, Nankai University, Tianjin 300071, China}

\begin{abstract}
A many-body quantum system with varying parameters can exhibit two distinct
quantum states within the same energy shell. This allows for a dynamic
transition from the ground state of the pre-quench Hamiltonian to a steady
state of the post-quench Hamiltonian. We investigate the dynamic response of
the ground states in a two-layer half-filled Hubbard model to a
perpendicular electric field. We demonstrate that the steady state exhibits
conductivity when the field is in resonance with the on-site repulsion,
while the initial state is a Mott-insulating state. Additionally, the two
layers exhibit identical conducting behavior due to the formation of
long-lived dopings, as evidenced by the charge fluctuation. The key factor
in achieving this dynamic transition is the cooperative interplay between
on-site interactions and the resonant field, rather than the individual
roles they play. Our findings offer an alternative mechanism for
field-induced conductivity in strongly correlated systems.
\end{abstract}

\maketitle

\section{Introduction}

Understanding the static and dynamical properties of strongly correlated
quantum systems poses a major challenge in both theory and experiment \cite%
{Dagotto1994,Georges1996,Yeshurun1996,Dukelsky2004,Basov2005,Bloch2008,Aoki2014}%
. Although the presence of strong quantum correlations often precludes
intuitive models, the interplay between particle-particle interaction and
kinetic energy gives rise to novel effects including the magnetism and
superconductivity \cite{Auerbach1994,Sierra1997,Giamarchi2016}. The Hubbard
model provides a minimal paradigm for studying these properties and, as
such, is attracting the attention of theoreticians from many different
corners of condensed matter \cite%
{Hubbard1964,Fisher1989,Yang1989,Georges1996,Lee2006,Sachdev2011,Keimer2015,Zhang2021a,Xie2023}%
. Recent years have seen a significant progress in simulating quantum
many-body physics with ultracold atoms that are well isolated from the
environment and therefore evolve under their intrinsic quantum dynamics \cite%
{Giorgini2008,Bloch2008,Randeria2012,Christian2017}. In particular,
ultracold fermionic atoms in an optical lattice can be used to realize the
Hubbard model \cite%
{Koehl2005,Joerdens2008,Schneider2008,Esslinger2010,Taie2012,Hart2015,Duarte2015,Cocchi2016,Hartke2020}%
. However, most real materials possess rather complex lattice structures,
which can be approximated as a system of coupled layers. Fortunately, a
bilayer Fermi--Hubbard model can be realized through a highly stable
vertical superlattice \cite%
{Gonzalez-Tudela2019,Koepsell2020,Gall2021,Meng2021}, which provides a
platform to study the basic property and enlarge our understanding of the
multi-layer interacting system.

Driven by the advances of high-resolution microscopy, the theory describing
the various quantum phase transitions has advanced remarkably in the context
of interacting many-body quantum systems at equilibrium. However, the
behavior of such systems is far less understood when it comes to the
non-equilibrium regime, whose relevance has rapidly grown triggered by the
significant experimental progress. Examples of systems that offer a large
degree of control such as ultracold atoms in optical lattices \cite%
{Greiner2002,Bloch2008,Esteve2008,Schneider2008,BakrW2010,Abanin2019},
trapped ions \cite{Albiez2005,Schumm2005,Hofferberth2007,BakrW2010,Blatt2012}%
, superconducting qubits \cite{Roushan2017,Xu2018}, as well as nuclear and
electron spins associated with impurity atoms in diamond \cite%
{Doherty2013,Schirhagl2013}. The tunability and long coherence times of
these systems, along with the ability to prepare highly non-equilibrium
states, enable one to probe quantum dynamics and thermalization in closed
systems \cite{Abanin2019}. Hence, it seems very timely to investigate the
non-equilibrium behavior of the bilayer Hubbard system and unravel the
potential exotic behavior.

The bilayer Hubbard model served as a paradigm to study band to Mott
insulator crossover \cite{Kancharla2007,Gall2021}. In the large $U$ limit,
whether the ground state is in a planar antiferromagnetically ordered Mott
insulating phase or a band insulating phase is fully determined by the ratio
of the inter-and intra-layer coupling. But no matter what phase the system
is in, its ground state is non-conductive. In this work, we propose a way to
realize the conductivity of the two-layer system through non-equilibrium
dynamics. To this end, we first prepare the system into a Mott insulator
phase with a strong antiferromagnetic correlation. Then we switch on a
resonant perpendicular electric field such that the single-occupied Mott
insulating states and non-ferromagnetic state doped with doublons and holes
hybridize in the same energy shell. This guarantees the significant charge
fluctuation of the low-energy eigenstates of the post-quench Hamiltonian
since the conducting proportion overwhelms the insulating ratio.
Furthermore, these states reside in the low-lying sector protected by the
energy gap. This keeps the two layers with dopings for a long time providing
a platform to facilitate the mobility of the particles in each layer under
the non-ferromagnetic background. Accordingly, the evolved state driven by
the post-quench Hamiltonian can hardly return to the initial insulating
state with zero charge fluctuation. Correspondingly, the system enters into
a non-equilibrium conducting phase featured by the large charge fluctuation.
The key ingredient of our proposal is the formation of the long-lived
dopings in each layer arising from the cooperation between the on-site
interaction and the resonant electric field. It is hoped that these results
can inspire further studies on the non-equilibrium bilayer even multi-layer
interacting system.

The remainder of the paper is structured as follows: In Sec. \ref{Model}, we
introduce the bilayer Hubbard model subjected to a perpendicular electric
field. Aiding by the symmetries of the considered system, the role of the
external field is elucidated in the absence of the on-site interaction. Sec. %
\ref{RF} is devoted to investigating the formation of the possible energy
shell that is determined by the cooperation between the resonant field and
on-site interaction. In Sec. \ref{carrier}, the mechanism of the
field-induced conducting state is sketched in which the long-lived dopings
play the pivotal role. Then a non-equilibrium scheme for generating a steady
conducting state is proposed. Finally, we summarize our results in In Sec. %
\ref{summary}.

\section{Two-layer Hubbard Model}

\label{Model}
\begin{figure}[tbp]
\centering
\includegraphics[width=0.8\textwidth]{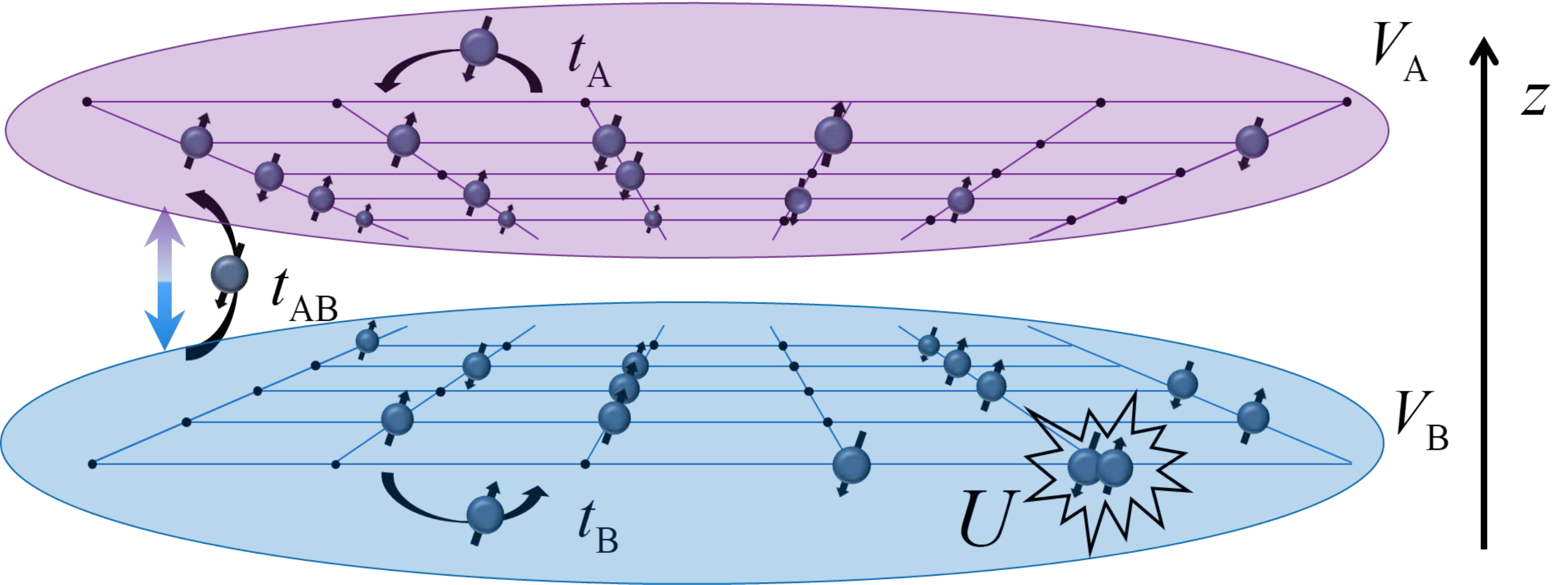}
\caption{Schematic illustration of the bilayer-Hubbard system subjected to a
perpendicular electric field. Experimentally, the bilayer system is realized
by a bichromatic superlattice in the $z$ direction trapping atoms in several
bilayer sheets \protect\cite{Gall2021}. The intra-layer hoppings $t_{\mathrm{%
A}}$ and $t_{\mathrm{B}}$ between lattice sites are controlled by the depth
of the optical lattices; the inter-layer tunnelling $t_{\mathrm{AB}}$, on
the other hand, is independently induced by microwaves \protect\cite%
{Meng2021}. Using tomographic imaging, (spin-) densities in a single layer
are detected.}
\label{fig_system}
\end{figure}
We consider a two-layer Hubbard model system, described by the Hamiltonian%
\begin{equation}
H=H_{\mathrm{A}}+H_{\mathrm{B}}+H_{\mathrm{AB}},  \label{H}
\end{equation}%
with sub-Hamiltonian on each layer%
\begin{eqnarray}
H_{\mathrm{A}} &=&t_{\mathrm{A}}\sum_{\sigma =\uparrow ,\downarrow
}\sum_{\left\langle i,j\right\rangle }(a_{i,\sigma }^{\dag }a_{j,\sigma }+%
\mathrm{H.c.})+U\sum_{i=1}^{N}a_{i,\uparrow }^{\dag }a_{i,\downarrow }^{\dag
}a_{i,\downarrow }a_{i,\uparrow }  \notag \\
&&+V_{\mathrm{A}}\sum_{i=1}^{N}\left( a_{i,\uparrow }^{\dag }a_{i,\uparrow
}+a_{i,\downarrow }^{\dag }a_{i,\downarrow }\right) ,  \notag \\
H_{\mathrm{B}} &=&t_{\mathrm{B}}\sum_{\sigma =\uparrow ,\downarrow
}\sum_{\left\langle i,j\right\rangle }(b_{i,\sigma }^{\dag }b_{j,\sigma }+%
\mathrm{H.c.})+U\sum_{i=1}^{N}b_{i,\uparrow }^{\dag }b_{i,\downarrow }^{\dag
}b_{i,\downarrow }b_{i,\uparrow }  \notag \\
&&+V_{\mathrm{B}}\sum_{i=1}^{N}\left( b_{i,\uparrow }^{\dag }b_{i,\uparrow
}+b_{i,\downarrow }^{\dag }b_{i,\downarrow }\right) ,
\end{eqnarray}%
and inter-layer hopping term%
\begin{equation}
H_{\mathrm{AB}}=t_{\mathrm{AB}}\sum_{\sigma =\uparrow ,\downarrow
}\sum_{j=1}^{L}(b_{j,\sigma }^{\dag }a_{j,\sigma }+\mathrm{H.c.}),
\end{equation}%
where $a_{i,\sigma }$\ and $b_{j,\sigma }$\ are fermion operators with spin-$%
\frac{1}{2}$ polarization $\sigma =\uparrow ,\downarrow $\ in layers $A$\
and $B$, respectively. The parameters $t_{\mathrm{A}}(t_{\mathrm{B}})$ and $%
t_{\mathrm{AB}}$\ are intra- and inter-layer hopping strengths, and taken to
be real in this paper. The on-site interaction $U$ tends to localize the
fermions by establishing a Mott insulator at half band filling. Here $H_{%
\mathrm{A}}$\ and $H_{\mathrm{B}}$ are identical, describing a standard
Hubbard model, which is restricted as simple bipartite lattice with no
restriction on their geometries. In particular, the key features of the
setup are: (i) $H_{\mathrm{AB}}$\ is the rung of the whole system\
representing the tunneling between two subsystems\ $H_{\mathrm{A}}$\ and $H_{%
\mathrm{B}}$. (ii) The on-site potentials $V_{\mathrm{A}}$ and $V_{\mathrm{B}%
}$. The gradient of such two uniform potentials acts as the electric field
that can bring the migration of particles. Such two ingredients play a
pivotal role in the subsequent quench dynamics. The schematic of the system
is presented in Fig. \ref{fig_system}. Driven by the advances\ of
high-resolution microscopy, such a bilayer Fermi--Hubbard model can be
realized experimentally by using ultracold atoms in an optical lattice. In
particular, the bichromatic optical superlattice in the vertical direction
can be employed to control the bilayer systems and inter-layer tunneling $t_{%
\mathrm{AB}}$ \cite{Koepsell2020,Hartke2020,Gall2021}. The other Hamiltonian
parameters $t_{\mathrm{A}}$, $t_{\mathrm{B}}$, $V_{\mathrm{A}}$, $V_{\mathrm{%
B}}$, $U$, the geometry \cite%
{Soltan-Panahi2011,Wirth2011,Tarruell2012,Jo2012,Shintaro2015}, and the
dimensionality are experimentally tunable, thus providing access to a large
parameter range.

The Hamiltonian in Eq. (\ref{H}) has a rich structure due to the symmetry it
possesses. To gain further insight into this model, we define pseudo-spin
operators%
\begin{eqnarray}
s^{+} &=&\left( s^{-}\right) ^{\dag }=\sum_{i=1}^{N}\left( a_{i,\uparrow
}^{\dag }a_{i,\downarrow }+b_{i,\uparrow }^{\dag }b_{i,\downarrow }\right) ,
\\
s^{z} &=&\frac{1}{2}\sum_{i=1}^{N}\left( a_{i,\uparrow }^{\dag
}a_{i,\uparrow }+b_{i,\uparrow }^{\dag }b_{i,\uparrow }-a_{i,\downarrow
}^{\dag }a_{i,\downarrow }-b_{i,\downarrow }^{\dag }b_{i,\downarrow }\right)
,
\end{eqnarray}%
which satisfy the Lie algebra commutation relations $\left[ s^{\mu },s^{\nu }%
\right] =\sum_{\lambda =x,y,z}2i\epsilon ^{\mu \nu \lambda }s^{\lambda }$,
and\ are employed to characterize the symmetry of the system. It can be
shown that the Hamiltonian has SU(2) symmetry, obeying%
\begin{equation}
\left[ H,s^{z}\right] =\left[ H,s^{\pm }\right] =0.
\end{equation}%
It has been shown that the ground state of $H$ with zero field at
half-filling is singlet with $s=0$ \cite{Lieb1989}, or anti-ferromagnetic
insulating state. The presence of the on-site potentials $V_{\mathrm{A}}$
and $V_{\mathrm{B}}$\ does not break the SU(2) symmetry. It has two
implications: (i) The appearance of the potential brings the possibility to
change the property of the original system, such as improving the mobility
of particle. (ii) The retained spin\ symmetry maintains the invariant
subspaces, which allows us to investigate the problem in each sector. And
every sector can be selected by adding an electric field in the experiment.

\begin{figure*}[tbp]
\centering
\includegraphics[width=0.9\textwidth]{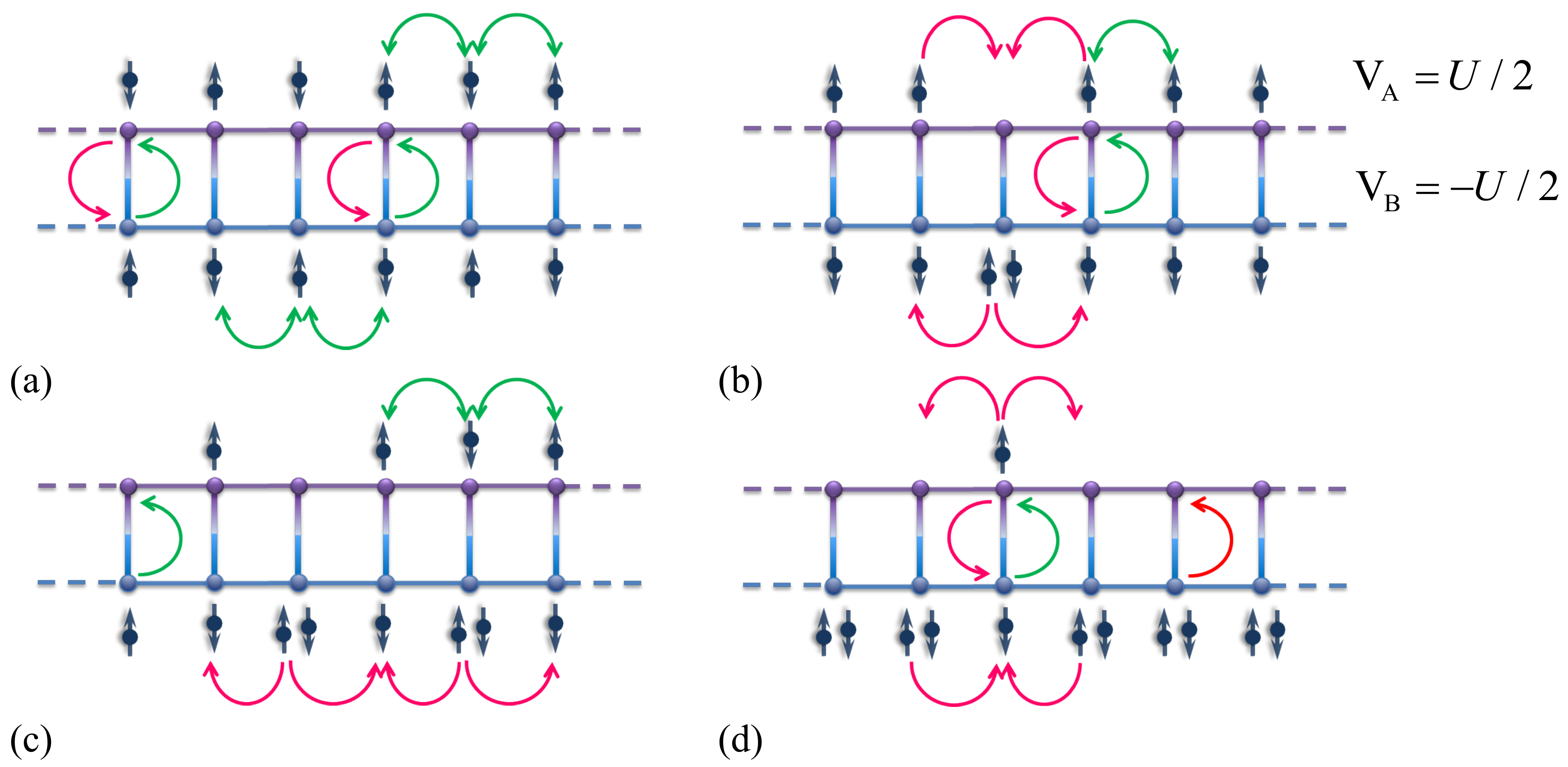}
\caption{Sketch of the possible movement of the particles driven by a
resonant electric field in a ladder system. Here we present several typical
configurations to show the underlying mechanism of the transition. (a) The
system is initialized in a single-occupied state such that the cooperation
between the resonant electric field $V_{\mathrm{A}}=-V_{\mathrm{B}}=U/2$ and
interaction $U$ only allows the migration of the particles from the $\mathrm{%
A}$ layer to the $\mathrm{B}$ layer. As a consequence, a hole is left in the
$\mathrm{A}$ layer while a doublon is formed in the $\mathrm{B}$ layer. But
the hopping of particles within each layer is strictly prohibited. (b)-(d)
denote the multi-doublon and hole cases. With the constraint of energy
conservation, the electron and hole can move in the non-ferromagnetic
background. In particular, the spin-up particle can move freely in the upper
layer while the spin-down particle (doublon) can effectively move as a
single particle with the aid of the background spin configuration. The
principle of the particle migration is that doublon cannot appear in the $%
\mathrm{A}$ layer and holes cannot appear in the $\mathrm{B}$ layer. The
hopping processes denoted by red arrows are allowed, while green arrows are
forbidden.}
\label{fig_illu}
\end{figure*}

Based on the aforementioned statement, we examine the response of the system
to the quenched perpendicular electric field when the interaction is free
for simplicity. To this end, we introduce the Fourier transformation in two
sub-systems,%
\begin{eqnarray}
a_{j,\sigma } &=&\frac{1}{\sqrt{N}}\sum_{k}e^{ikj}a_{k,\sigma }, \\
b_{j,\sigma } &=&\frac{1}{\sqrt{N}}\sum_{k}e^{ikj}b_{k,\sigma },
\end{eqnarray}%
where the wave vector $k=2n\pi /N$, $\left( n\in \left[ -L/2,\text{ }L/2-1%
\right] \right) $. This transformation block diagonalizes the Hamiltonian
due to its translational symmetry, i.e.,
\begin{equation}
H=\sum_{k,\sigma }H_{k,\sigma }=\sum_{k,\sigma }\psi _{k,\sigma }^{\dagger
}h_{k,\sigma }\psi _{k,\sigma },
\end{equation}%
satisfying $\left[ H_{k,\sigma },\text{ }H_{k^{\prime },\sigma ^{\prime }}%
\right] =0$. Here $H$ is rewritten in the Nambu representation with the basis%
\begin{equation}
\psi _{k,\sigma }=\left(
\begin{array}{c}
a_{k,\sigma } \\
b_{k,\sigma }%
\end{array}%
\right) ,
\end{equation}%
and $h_{k,\sigma }$ is a core matrix with the form
\begin{equation}
h_{k,\sigma }=\left(
\begin{array}{cc}
\lambda _{\mathrm{A}} & t_{\mathrm{AB}} \\
t_{\mathrm{AB}} & \lambda _{\mathrm{B}}%
\end{array}%
\right) ,
\end{equation}%
where $\lambda _{\mathrm{\sigma }}=-2t_{\parallel }\cos k+V_{\mathrm{\sigma }%
}$ ($\mathrm{\sigma =A,}$ $\mathrm{B}$). Note that $t_{\mathrm{A}}=t_{%
\mathrm{B}}=t_{\parallel }$ is assumed for simplicity. The Hamiltonian can\
be diagonalized by taking the linear transformation on $h_{k,\sigma }$.
Straightforward algebra shows that
\begin{equation}
H=\sum_{k,\sigma }\left( \zeta _{k}+\varepsilon \right) \alpha _{k,\sigma
}^{\dagger }\alpha _{k,\sigma }+\left( \zeta _{k}-\varepsilon \right) \beta
_{k,\sigma }^{\dagger }\beta _{k,\sigma },
\end{equation}%
where%
\begin{eqnarray}
\zeta _{k} &=&-2t_{\parallel }\cos k+\frac{V_{\mathrm{A}}+V_{\mathrm{B}}}{2},
\\
\varepsilon &=&\sqrt{t_{\mathrm{AB}}^{2}+\left( \frac{V_{\mathrm{A}}-V_{%
\mathrm{B}}}{2}\right) ^{2}},
\end{eqnarray}%
and
\begin{eqnarray}
\alpha _{k,\sigma }^{\dagger } &=&\cos \frac{\theta }{2}a_{k,\sigma
}^{\dagger }+\sin \frac{\theta }{2}b_{k,\sigma }^{\dagger }, \\
\beta _{k,\sigma }^{\dagger } &=&\sin \frac{\theta }{2}a_{k,\sigma
}^{\dagger }-\cos \frac{\theta }{2}b_{k,\sigma }^{\dagger },
\end{eqnarray}%
with $\cos \theta =\left( V_{\mathrm{A}}-V_{\mathrm{B}}\right) /2\varepsilon
$. Evidently, the role of the tilted potential and tunneling $t_{\mathrm{AB}%
} $ is to separate two identical single-particle energy bands. Hence, the
ground state of the system is the direct product of spin-singlets and
triplets belonging to the different $k$ subspace, given that the influence
of the electric field is not considered and the system is at half-filling.
The corresponding ratio between such two components is determined by $t_{%
\mathrm{AB}}$. In particular, the ground state is a band insulator of
spin-singlets along the bonds between the layers when $t_{\mathrm{AB}%
}/t_{\parallel }>2$. Without loss of generality, we take the ground state of
$H$ with $V_{\mathrm{A}}=V_{\mathrm{B}}=0$ at half-filling as the initial
state
\begin{equation}
\left\vert \psi \left( 0\right) \right\rangle =\prod_{\left\vert k^{\prime
}\right\vert <k_{c}^{\prime },\sigma }a_{k^{\prime },\sigma }^{\dagger
}b_{k^{\prime },\sigma }^{\dagger }\prod_{k_{c}^{\prime }<\left\vert
k\right\vert <k_{c},\sigma }\frac{a_{k,\sigma }^{\dagger }-b_{k,\sigma
}^{\dagger }}{\sqrt{2}}\left\vert \mathrm{Vac}\right\rangle ,
\end{equation}%
where $\left\vert \mathrm{Vac}\right\rangle $ is the vacuum state of
fermion. Here $k_{c}^{\prime }=$\textrm{min}$\{$\textrm{arccos}$(t_{\mathrm{%
AB}}/2t_{\parallel })$, \textrm{arccos}$(-t_{\mathrm{AB}}/2t_{\parallel })\}$%
, and $k_{c}=\pi -k_{c}^{\prime }$. After a quench, the on-site potentials $%
V_{\mathrm{A}}$ and $V_{\mathrm{B}}$ are applied, then the evolve state
driven by $H$ can be given as
\begin{eqnarray}
\left\vert \psi \left( t\right) \right\rangle &=&\prod_{\left\vert k^{\prime
}\right\vert <k_{c}^{\prime },\sigma }e^{-2i\zeta _{k^{\prime
}}t}a_{k^{\prime },\sigma }^{\dagger }b_{k^{\prime },\sigma }^{\dagger
}\times  \notag \\
&&\prod_{k_{c}^{\prime }<\left\vert k\right\vert <k_{c},\sigma }e^{-i\zeta
_{k}t}\frac{P_{-}a_{k,\sigma }^{\dagger }-P_{+}b_{k,\sigma }^{\dagger }}{%
\sqrt{2}}\left\vert \mathrm{Vac}\right\rangle ,
\end{eqnarray}%
where $P_{\pm }=\cos \varepsilon t+i\sin \varepsilon t\left( \sin \theta \pm
\cos \theta \right) $. To measure the migration of particles, the ratio of
the particle density between two layers
\begin{equation}
r(t)=\frac{\langle \sum_{\sigma ,i=1}^{N}a_{i,\sigma }^{\dag }a_{i,\sigma
}\rangle }{\langle \sum_{\sigma ,i=1}^{N}b_{i,\sigma }^{\dag }b_{i,\sigma
}\rangle },
\end{equation}%
is introduced. For the evolved state $\left\vert \psi \left( t\right)
\right\rangle $, its transfer rate can be given as
\begin{equation}
r(t)=\frac{\sum_{\left\vert k^{\prime }\right\vert <k_{c}^{\prime },\sigma
}1+\frac{1}{2}\sum_{k_{c}^{\prime }<\left\vert k\right\vert <k_{c},\sigma }%
\left[ 1-\sin 2\theta \sin ^{2}\varepsilon t\right] }{\sum_{\left\vert
k^{\prime }\right\vert <k_{c}^{\prime },\sigma }1+\frac{1}{2}%
\sum_{k_{c}^{\prime }<\left\vert k\right\vert <k_{c},\sigma }\left[ 1+\sin
2\theta \sin ^{2}\varepsilon t\right] }.
\end{equation}%
Evidently, $r(t)$ exhibits a periodical behavior with period $\tau =\pi
/\varepsilon $. Further, the averaged transfer ratio over a long time
interval $T$ is defined as
\begin{equation}
\overline{r}=\lim_{T\rightarrow \infty }\frac{1}{T}\int_{0}^{T}r(t)\mathrm{d}%
t.
\end{equation}%
Direct derivation shows that%
\begin{equation}
\overline{r}=\frac{\sum_{\left\vert k^{\prime }\right\vert <k_{c}^{\prime
},\sigma }1+\frac{1}{2}\sum_{k_{c}^{\prime }<\left\vert k\right\vert
<k_{c},\sigma }\left[ 1-\sin \theta \cos \theta \right] }{\sum_{\left\vert
k^{\prime }\right\vert <k_{c}^{\prime },\sigma }1+\frac{1}{2}%
\sum_{k_{c}^{\prime }<\left\vert k\right\vert <k_{c},\sigma }\left[ 1+\sin
\theta \cos \theta \right] }.  \label{average_r}
\end{equation}%
It indicates that the cooperation of the perpendicular field and $t_{\mathrm{%
AB}}$ leads to the transfer of particles between the different layers, which
serves as the building block to realize the conducting phase from a Mott
insulator as $U$ switches on. Note, in passing, that the direction of
particle transfer depends on the choice of the initial state and is not
necessarily related to the direction of gradient of the chemical potential.
An obvious example is the sign reversal of $\theta $ in Eq. (\ref{average_r}%
) for the highest excited state, where particles tend to go from layer $%
\mathrm{B}$ to layer \textrm{A. }It is in stark difference from the
classical system in which the flow of particles is along the direction of
the gradient of the potential. On the other hand, in the absence of $U$, the
transfer rate between two layers oscillates such that the system acquires
the information of the initial state once the time $t=n\tau $ ($n$ is a
positive integer). In other words, the presence of potential only induces
the particle migration and does not result in the imbalance of particle
density between two layers over a long time interval.

In the following, we study the dynamical response of the ground state of $H$
(large $U$) to a quenched electric field with the resonant value.

\begin{figure*}[tbp]
\centering
\includegraphics[width=0.9\textwidth]{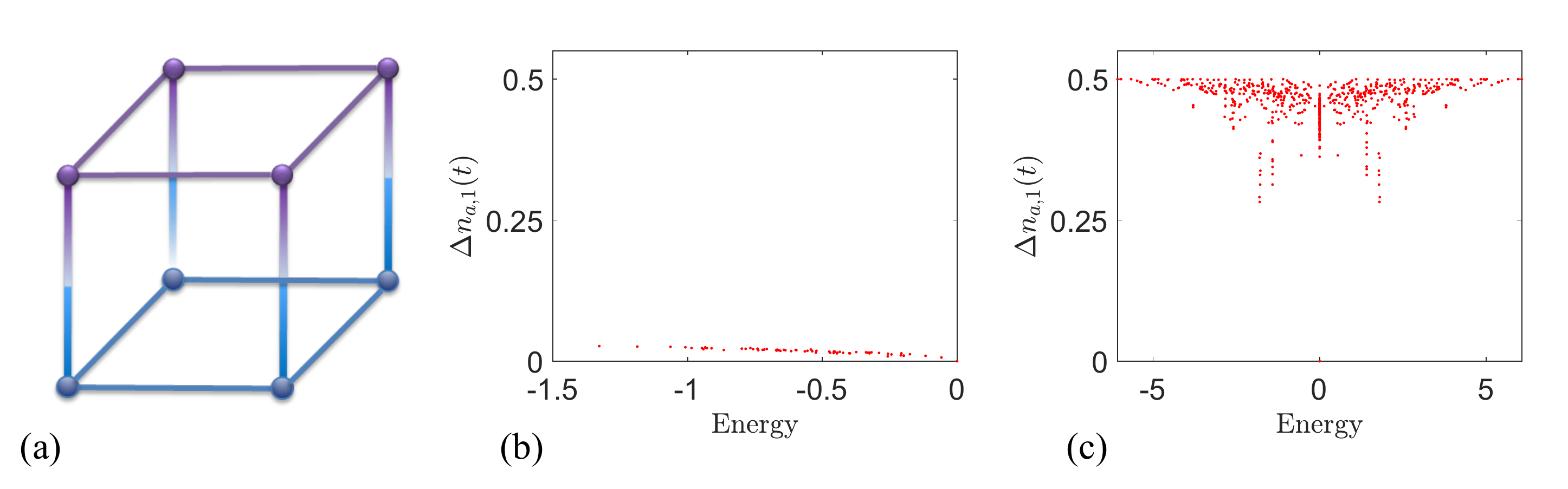}
\caption{Plots of the local charge fluctuation $\Delta n_{a,1}$ for the low
energy excitation of the Hubbard model at half-filling. The geometry of the
considered two-layer system is illustrated in Fig. \protect\ref{fig1}(a).
The other system parameters are given as (b) $t_{\mathrm{B}}/t_{\mathrm{A}%
}=1 $, $V=0$, and $U/t_{\mathrm{A}}=20$ (c) $t_{\mathrm{B}}/t_{\mathrm{A}}=1$%
, $V/t_{\mathrm{A}}=10$, and $U/t_{\mathrm{A}}=20$. In Fig. \protect\ref%
{fig1}(b), the system favors the Mott-insulating phase and hence $\Delta
n_{a,1}$ of low-energy eigenstates are around $0$. However, $\Delta n_{a,1}$
is around $0.5$ except for very few eigenstates when the resonant field $V$
presents. The cooperation between $U$ and $V$ results in a dramatic change
in the system conductivity.}
\label{fig1}
\end{figure*}

\section{Resonant field}

\label{RF}

Meanfield theory \cite{Sachdev2011} has shown that the phase diagram of a
Hubbard model is described by the Mott lobe, which is a function of $U$ and
chemical potential. The formation of the anti-ferromagnetic insulating
ground state requires large $U$ and half-filling. For the bilayer system, we
note that the total particle number of the two-layer is conservative, while
the one of each layer is not. In this work we consider the case with fixed
particle number, which equals to $2N$. Thus when taking $V_{\mathrm{A}}=V_{%
\mathrm{B}}$,\ the ground state is an anti-ferromagnetic insulating state.
However, when taking $V_{\mathrm{A}}\neq V_{\mathrm{B}}$, the chemical
potentials can adjust the particle density on each layer, since one layer
can act as a source of particles for the other. It happens by the difference
of the chemical potential, or the perpendicular electric field. In the
following, we will show that an optimal field can provide a resonant channel
for the directional migration of particles from one layer to the other.

We start with the simplest case with $t_{\mathrm{A}}=t_{\mathrm{B}}=t_{%
\mathrm{AB}}=V_{\mathrm{A}}=V_{\mathrm{B}}=0$. The Hamiltonian only contains
the repulsive term, i.e., $U\sum_{i=1}^{N}\left( a_{i,\uparrow }^{\dag
}a_{i,\downarrow }^{\dag }a_{i,\downarrow }a_{i,\uparrow }+b_{i,\uparrow
}^{\dag }b_{i,\downarrow }^{\dag }b_{i,\downarrow }b_{i,\uparrow }\right) $,
and then the ground states in the subspace with zero $s^{z}$\ are\
multi-degenerate. The energy is zero and the degenerate degree is $\mathcal{N%
}_{\mathrm{I}}=C_{2N}^{N}$, the explicit expression of which is $%
C_{2N}^{N}=(2N)!/[(2N-N)!(N)!]$. These states construct singlet ground state
when parameters $t_{\mathrm{A}}=t_{\mathrm{B}}=t_{\mathrm{AB}}$ are switched
on in large $U$ limit, and are referred to as the insulating set (I-set) of
states in this paper.

Now we consider the case with $t_{\mathrm{A}}=t_{\mathrm{B}}=t_{\mathrm{AB}%
}=0$, and $V_{\mathrm{A}}=-V_{\mathrm{B}}=U/2$ (hereafter $V_{\mathrm{A}%
}=-V_{\mathrm{B}}=V$), which is referred to as the resonant field. The
Hamiltonian can be rewritten in the from%
\begin{eqnarray}
H_{\mathrm{A}} &=&U/2\sum_{i=1}^{N}\left( a_{i,\uparrow }^{\dag
}a_{i,\uparrow }+a_{i,\downarrow }^{\dag }a_{i,\downarrow }\right) ^{2}, \\
H_{\mathrm{B}} &=&-U/2\sum_{i=1}^{N}\left( b_{i,\uparrow }^{\dag
}b_{i,\uparrow }-b_{i,\downarrow }^{\dag }b_{i,\downarrow }\right) ^{2},
\end{eqnarray}%
which is convenient for the following analysis. The eigen energy of $H_{%
\mathrm{A}}+H_{\mathrm{B}}$\ is%
\begin{equation}
E=U/2\sum_{i=1}^{N}\left[ \left( n_{\mathrm{A,}i,\uparrow }+n_{\mathrm{A,}%
i,\downarrow }\right) ^{2}-\left( n_{\mathrm{B,}i,\uparrow }-n_{\mathrm{B,}%
i,\downarrow }\right) ^{2}\right] ,
\end{equation}%
where $n_{a\mathrm{,}i,\sigma }$ and $n_{b\mathrm{,}i,\sigma }$\ are eigen
values of operators $a_{i,\sigma }^{\dag }a_{i,\sigma }$\ and $b_{i,\sigma
}^{\dag }b_{i,\sigma }$, respectively. For the half-filled case in the
subspace with zero spin component in $z$ direction, we have the following
constraint conditions
\begin{eqnarray}
\sum_{\sigma =\uparrow ,\downarrow }\left( N_{a\mathrm{,}\sigma }+N_{b%
\mathrm{,}\sigma }\right) &=&2N, \\
N_{a\mathrm{,}\uparrow }+N_{b\mathrm{,}\uparrow }-N_{a\mathrm{,}\downarrow
}-N_{b\mathrm{,}\downarrow } &=&0,
\end{eqnarray}%
\ where $N_{a\mathrm{,}\sigma }$ and $N_{b\mathrm{,}\sigma }$\ denote the
eigen values of operators $\sum_{i=1}^{N}a_{i,\sigma }^{\dag }a_{i,\sigma }$
and $\sum_{i=1}^{N}b_{i,\sigma }^{\dag }b_{i,\sigma }$, respectively. Notice
that when every site is occupied by a single particle, we have $\left( n_{a%
\mathrm{,}i,\uparrow }+n_{a\mathrm{,}i,\downarrow }\right) ^{2}=\left( n_{b%
\mathrm{,}i,\uparrow }-n_{b\mathrm{,}i,\downarrow }\right) ^{2}=1$ for all $%
i\in \left[ 1,N\right] $ such that the ground state energy is $E=0$. In the
following, we analyse all the possible eigenstates with zero energy, since
the number and configuration of which are important for our results. As the
mentioned above, the number of single-occupied states is $\mathcal{N}_{%
\mathrm{I}}$ and the superposition of these states constructs the
anti-ferromagnetic groundstate of the original Hamiltonian with large $U$
and zero $V$. Based on such states, losing a particle in layer $\mathrm{A}$
results in energy lose of $U/2$, while gaining a fermion in layer \textrm{B}
results in energy lose of $-U/2$, generating a new zero energy state.
Accordingly, a set of zero energy states can be constructed by losing $n$ ($%
n\in \left[ 1,N\right] $) particles in layer $\mathrm{A}$ and gaining $n$
particle in layer \textrm{B}. The number of such kind of states is $\left(
C_{N}^{n}\right) ^{2}C_{2N-2n}^{N-n}$, in which term $C_{N}^{n}$\ comes from
the configuration of hole in layer $\mathrm{A}$ or the doublon in layer
\textrm{B}, and $C_{2N-2n}^{N-n}$ comes from the spin configuration of the
rest single-occupied particles. Then the total number of zero-energy
(ground) states is $\mathcal{N}_{\mathrm{C}}=%
\sum_{j=1}^{N}(C_{N}^{j})^{2}C_{2N-2j}^{N-j}$. All these states are
hybridized by the inter-chain and intra-chain hoppings. In other words, the
hopping terms can drive one of these states to all the rest states.

\section{the mobility of the free doublons and holes}

\label{carrier} In this section, we discuss the observable results based on
the analysis in the last section. We can classify all states into two sets:
(i) \textrm{I}-set (insulating set), as mentioned above, which consists of
complete single-occupied states; (ii) \textrm{C}-set (conducting set), which
consists of states involving holes and doublons. The reason for such a
classification is that the superposition of states in \textrm{I}-set
describes an antiferromagnetic Mott insulating state, while the
superposition of states in \textrm{C}-set describes a conducting state. We
explain the reason for the classification via the following examples. For
simplicity, we consider the simplest case involving states with single-hole
and single-doublon. \textrm{A} set representative of states in \textrm{C}%
-set can be expressed as the form $a_{i,\uparrow }b_{i,\uparrow }^{\dag
}\left\vert \Uparrow \right\rangle _{\mathrm{A}}\left\vert \Downarrow
\right\rangle _{\mathrm{B}}$, where states $\left\vert \Uparrow
\right\rangle _{\mathrm{A}}=\prod_{l=1}^{N}a_{l,\uparrow }^{\dag }\left\vert
\mathrm{Vac}\right\rangle $\ and $\left\vert \Downarrow \right\rangle _{%
\mathrm{B}}=\prod_{l=1}^{N}b_{l,\downarrow }^{\dag }\left\vert \mathrm{Vac}%
\right\rangle $ denote ferromagnetic states of \textrm{A} and \textrm{B}
layers, respectively, and $\left\vert \mathrm{Vac}\right\rangle $\ is the
empty state of the whole system. Obviously, the dynamics of hole and
electron on each layers can act as free carriers on the ferromagnetic
background. Likewise, similar situation occurs for multi-hole and doublon
cases, and for other types of non-ferromagnetic background. For clarity, we
sketch some typical cases in Fig. \ref{fig_illu}.
\begin{figure*}[tbp]
\centering
\includegraphics[width=0.8\textwidth]{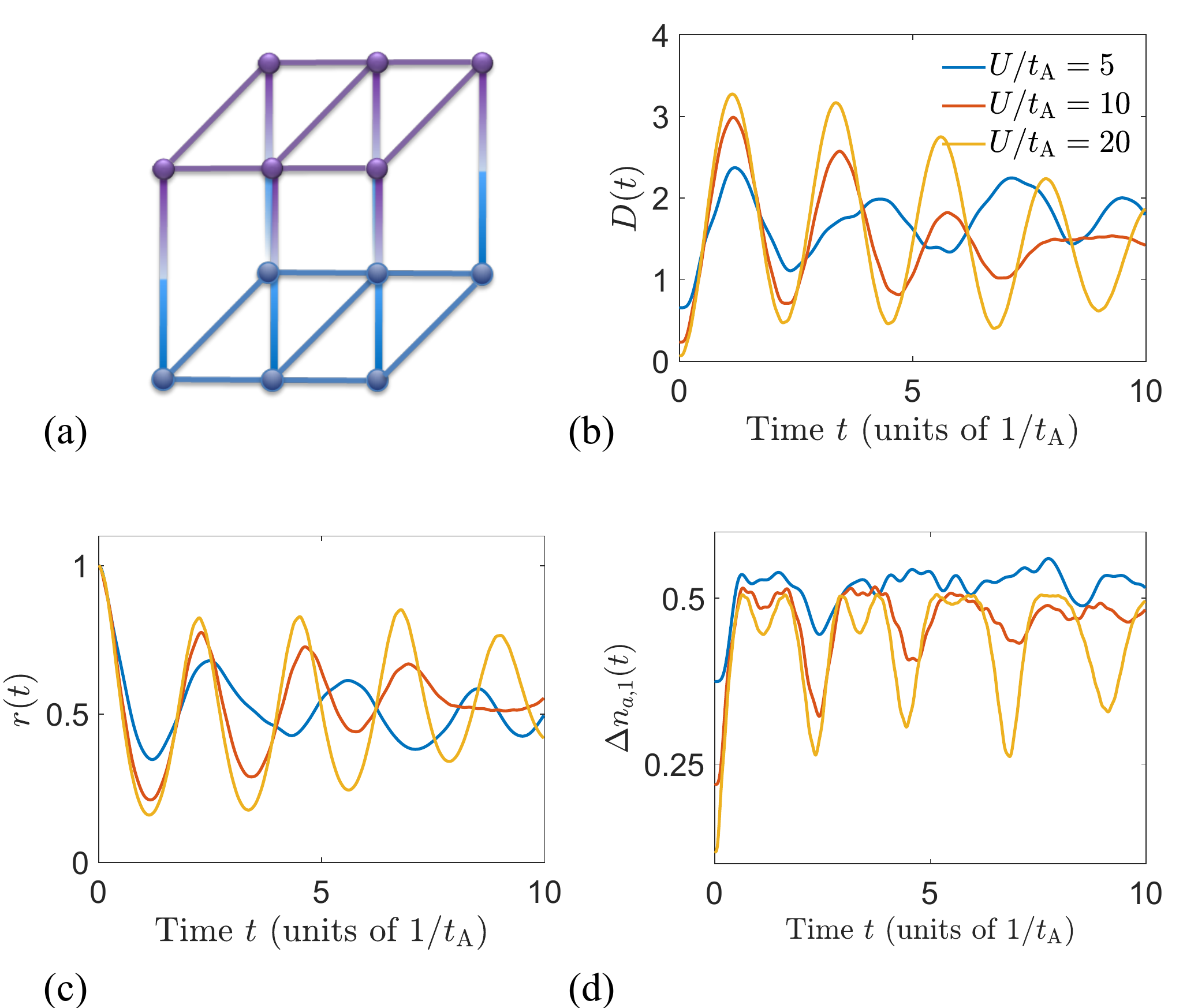}
\caption{Numerical simulation of the quenched bilayer Hubbard system for
different repulsive interactions $U/t_{\mathrm{A}}=5$, $10$, and $20$. The
system consists of two coupled ladders and is initialized in the ground
state of $H_{\mathrm{pre}}$. The small charge fluctuation of the initial
ground state arises from the finite interaction which allows the existence
of a double-occupied state with $s=0$. Such non-ferromagnetic kind of states
is suppressed when $U$ increases. After a quench, the doublons and holes
appear in the evolved state due to the particle flow from the upper layer to
the lower layer as shown in Fig. \protect\ref{fig1}(a). This is protected by
the energy shell residing in the low-energy excited subspace. In the
non-ferromagnetic background, the non-equilibrium system enters into the
conducting phase manifested by the increase of the charge fluctuation. Note
that the layers \textrm{A} and \textrm{B} share the same charge fluctuation $%
\Delta n_{a,j}=\Delta n_{b,j}=\Delta n_{j}$ since both free electrons and
holes play the same role in the particle migration. }
\label{fig2}
\end{figure*}
In order to predict the dynamical behavior of the quench, we estimate the
following two quantities. The first one is the ratio of the dimensions of
insulating and conducting classes. Straightforward calculation gives%
\begin{equation}
r_{1}=\frac{\mathcal{N}_{\mathrm{I}}}{\mathcal{N}_{\mathrm{C}}}\approx
e^{-\gamma N},
\end{equation}%
where factor $\gamma =0.325$\ is obtained from the numerical fitting. The
second one is the ratio of the average particle numbers between insulating
and conducting classes. Considering a mixed state, in which all the
zero-energy states have the same probability, we have the average numbers on
each layers%
\begin{eqnarray}
\overline{N}_{\mathrm{A,}\uparrow } &=&\overline{N}_{\mathrm{A,}\downarrow }=%
\frac{\sum_{j=1}^{N}j(C_{N}^{j})^{2}C_{2N-2j}^{N-j}}{\mathcal{N}_{\mathrm{I}%
}+\mathcal{N}_{\mathrm{C}}}=\frac{N}{3}, \\
\overline{N}_{\mathrm{B,}\uparrow } &=&\overline{N}_{\mathrm{B,}\downarrow }=%
\frac{\sum_{j=1}^{N}j(C_{N}^{j})^{2}C_{2j}^{j}}{\mathcal{N}_{\mathrm{I}}+%
\mathcal{N}_{\mathrm{C}}}=\frac{2N}{3},
\end{eqnarray}%
which result in
\begin{equation}
r_{2}=\frac{\overline{N}_{\mathrm{A,}\uparrow }+\overline{N}_{\mathrm{A,}%
\downarrow }}{\overline{N}_{\mathrm{B,}\uparrow }+\overline{N}_{\mathrm{B,}%
\downarrow }}=1/2.
\end{equation}%
These two quantities imply the following predictions. (i) The initial state
has a very small portion ($r_{1}$) comparing to the possible evolved states,
which will never go back to the initial one. It is similar to the case with
a site initial state in a square lattice. The probability will spread out,
approaching to uniform distribution after a long time. Then the evolved
state probably relaxes to a conducting state. (ii) The value of $r_{2}$\
indicates that the ratio of numbers on two layers turns to be $0.5$ after a
long time. It predicts that there is a particle flow in the quench dynamical
process.

In order to capture the main consequence of the particle flow arising from
the cooperation of $U$ and resonant $V$, we further introduce the local
charge fluctuation $\Delta n_{\lambda ,j}$ to compare the conductivity of
the eigenstates before and after quench. The definition of $\Delta
n_{\lambda ,j} $ is given by

\begin{equation}
\Delta n_{\lambda ,j}=\sqrt{\Delta _{1,\lambda ,j}-\Delta _{2,\lambda ,j}^{2}%
}\quad(\lambda =a, b),
\end{equation}%
where
\begin{eqnarray}
\Delta _{1,\lambda ,j} &=&\langle (\sum_{\sigma }\lambda _{j,\sigma }^{\dag
}\lambda _{j,\sigma })^{2}\rangle , \\
\Delta _{2,\lambda ,j} &=&\langle \sum_{\sigma }\lambda _{j,\sigma }^{\dag
}\lambda _{j,\sigma }\rangle .
\end{eqnarray}%
Specifically, if the system is in the Mott insulating phase, such
fluctuation would become energetically unfavorable, forcing the system into
a number state and exhibiting a vanishing particle number fluctuation $%
\Delta n_{\lambda ,j}$. Beyond the Mott insulator regime, the fermions are
delocalized with the non-vanishing charge fluctuation. In Fig. \ref{fig1},
we figure out the conductivity of the eigenstates in the cases of $H\left(
V=0\right) $ and $H\left( V=U/2\right) $, respectively. The system is at
symmetric half filling, i.e., $s^{z}=0$, and its geometry is illustrated in
Fig. \ref{fig1}(a). We focus on the low-lying eigenstates whose energy is
around $0$. When $V=0$ and large $U$ limit, the low energy excitation of the
considered half-filled Hubbard model can be effectively described by the
Heisenberg model with two layers. The corresponding eigenstates are the
superposition of bases belonging to the I-set. As such, the charge
fluctuation $\Delta n_{a,1}$ is about $0$ as shown in Fig. \ref{fig1}(b).
Conversely, the low-lying eigenstates of $H\left( V=U/2\right) $ possess the
non-zero charge fluctuation since C-set dominates especially in the large $N$
limit. This can be seen in Fig. \ref{fig1}(c). The local charge fluctuation
for most eigenstates is $0.5$. To gain further insight into the dynamics of
the two-layer system, we consider an initial state expanded in such energy
shell,
\begin{equation}
\left\vert \psi \left( 0\right) \right\rangle =\sum_{\alpha }c_{\alpha
}\left\vert E_{\alpha }\right\rangle \text{,}
\end{equation}%
where $\left\vert E_{\alpha }\right\rangle $ is the eigenstate of $H$ with
eigenvalue $E_{\alpha }$ in the low-lying energy shell. The expectation
value of a local operator $\widehat{O}$ as a function of time thus reads
\begin{eqnarray}
O\left( t\right) &\equiv &\left\langle \psi \left( t\right) \right\vert
\widehat{O}\left\vert \psi \left( t\right) \right\rangle  \notag \\
&=&\sum_{\alpha }\left\vert c_{\alpha }\right\vert ^{2}O_{\alpha \alpha
}+\sum_{\alpha \neq \beta }O_{\alpha \beta }e^{i\left( E_{\alpha }-E_{\beta
}\right) t},  \label{O_a}
\end{eqnarray}%
where $O_{\alpha \alpha }=\left\langle E_{\alpha }\right\vert \widehat{O}%
\left\vert E_{\alpha }\right\rangle $ and $\widehat{O}$ stands for either
the operator $(\sum_{\sigma }\lambda _{j,\sigma }^{\dag }\lambda _{j,\sigma
})^{2}$ or $\sum_{\sigma }\lambda _{j,\sigma }^{\dag }\lambda _{j,\sigma }$
in the considered two-layer system. Due to phase cancellation of the
off-diagonal terms in Eq. (\ref{O_a}), the long-time average is determined
only by the average in the diagonal element such that
\begin{equation}
\overline{O}\left( t\right) =\lim_{T\rightarrow \infty }\frac{1}{T}%
\int_{0}^{T}O\left( t\right) \mathrm{d}t=\sum_{\alpha }\left\vert c_{\alpha
}\right\vert ^{2}O_{\alpha \alpha }.
\end{equation}%
For the considered two types of the systems $H\left( V=0\right) $ and $%
H\left( V=U/2\right) $, $O_{\alpha \alpha }$ approaches a constant number
for the most eigenstates of the low-lying energy sector. Hence, $\overline{O}%
\left( t\right) =O_{\alpha \alpha }$ indicating that the eigenstate and
arbitrary evolved state share the same average value of $\Delta n_{\lambda
,j}$. It can be expected that the\ initial state containing the insulating
component will tend to the conducting state after a long-time evolution. In
the subsequent section, we will demonstrate this point through a quench
process.

Before ending this section, we want to point out that the conducting
mechanism is different from that of the single-layer system wherein the
particles are not half-filling. For the latter system, the maximum of holes
is contingent on the filled particle number. As for the two-layer system,
layer \textrm{A} acts as a source of particles for layer \textrm{B}, and the
corresponding particle number of layer \textrm{A} is not conserved. As a
consequence, layer \textrm{A }involves all the possible configurations of
holes in the low-energy sector, which provides more channels for the
transfer of particles between the layers. The cooperation between $U$ and
the resonant $V$ can stabilize the system in this phase for a long time.

\section{Dynamical density transfer}

\label{transfer}

The above analysis provides a prediction about the dynamical transition from
an insulating state to a conducting state in a composite system. In a
slightly different language, the main point of the above consideration can
be viewed as follows. A Mott insulating state is essentially the result of
Pauli exclusion principle and Coulomb repulsion at half-filling. It is not a
stable equilibrium since the chemical potential can change the particle
density for an open system. For a half-filled bilayer Hubbard system with
zero fields, the ground state lies at the equilibrium point for each layer.
When the resonant external field is switched on, the low-energy eigenstates
changes dramatically with positive or negative doping in each layer. This
exactly compensates for repulsive interactions presents in the Hubbard model
such that the transfer of particles between the two layers cost free. As a
consequence, many nearly degenerate conducting states lie at the low-energy
sector of quenched Hamiltonian which allows an easier transfer of particles
between layers.

We perform numerical simulations on a finite system for the following quench
process. The pre-quench Hamiltonian is assumed as $H_{\mathrm{pre}}=H$ with $%
t_{\mathrm{A}}=t_{\mathrm{B}}=t_{\mathrm{AB}}=1$, and $V=0$. The post-quench
Hamiltonian is then $H_{\mathrm{pos}}=H$ with $V=U/2$. Initially, the system
is half-filled and at the ground state, being an insulating state with a
strong anti-ferromagnetic correlation. According to our analysis, the quench
dynamics should result in a steady conducting state, realizing the dynamical
transition from an insulating state to a conducting state. The evolved
states $\left\vert \Psi \left( t\right) \right\rangle $\ for initial
anti-ferromagnetic ground state $\left\vert \Psi \left( 0\right)
\right\rangle $ is calculated by exact diagonalization method. We focus on
the following quantities: (i) The ratio of particle density between two
layers $r(t)$ which is a key signature of the dynamical transition; (ii) The
number fluctuation $\Delta n_{\lambda ,j}$ of the evolved state $\left\vert
\Psi \left( t\right) \right\rangle $.
\begin{figure}[tbp]
\centering
\includegraphics[width=0.8\textwidth]{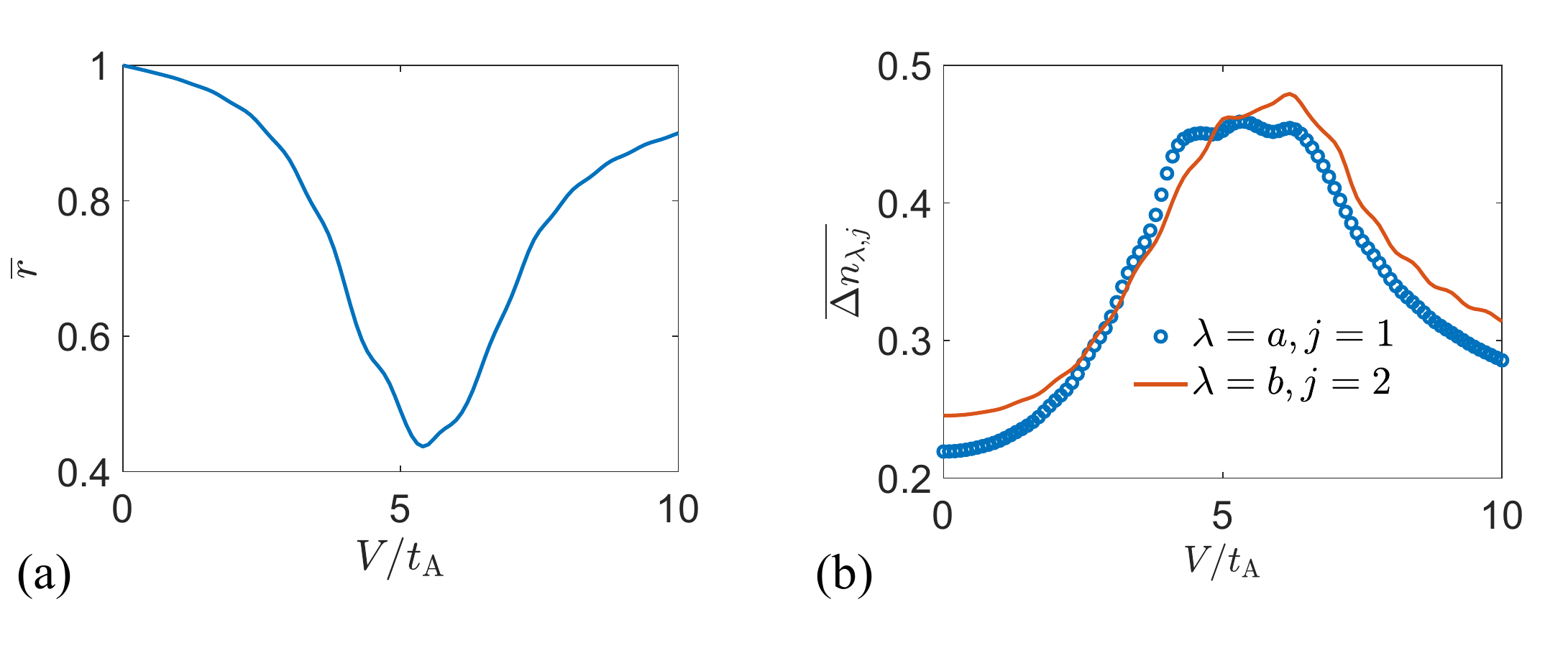}
\caption{Average $\overline{r}$ and $\overline{\Delta n_{\protect\lambda ,j}}
$ as functions of $V$ for $12$-site Hubbard model at half filling and $%
s_{z}=0$. The initial state is prepared as the ground state of $H_{\mathrm{%
pre}}$. The system parameters are $U/t_{\mathrm{A}}=10$ and cutoff quench
time $T=50/t_{\mathrm{A}}$. It indicates that, as $V$ approaches $U/2$, a
dip will appear verifying our prediction in the text.}
\label{fig3}
\end{figure}
The lattice geometry and numerical results are plotted in Fig. \ref{fig2},
where a $12$-site Hubbard model at half filling is considered. It can be
shown that the resonant electric field can induce the particle transfer from
the upper layer to the lower layer forming the doublons and holes in the
evolved state. This can be demonstrated by the increase of total double
occupancy $D=\sum_{j}\langle D_{j}\rangle $, where the local
double-occupation operator is given by $D_{\lambda ,j}=\lambda _{j,\uparrow
}^{\dagger }\lambda _{j,\uparrow }\lambda _{j,\downarrow }^{\dagger }\lambda
_{j,\downarrow }$ ($\lambda =a,$ $b$). The movable electrons and holes
significantly increase the number fluctuations of both layers and hence
enhances the conductivity. Such a mechanism still holds even in the small $U$
condition which can be shown by comparing Figs. \ref{fig2} (b)-(d). The weak
interaction means that the ground state does not fully consist of the
single-occupied Mott insulating state such that the initial double occupancy
and charge fluctuation is not zero, which can be shown in Figs. \ref{fig2}
(b) and (d). Instead, the portion of the other state $s=0$, i.e., doublon
state, increases leading to the non-zero number fluctuation of the initial
state and total transfer rate between the two layers. It indicates that the
effectiveness of the scheme is suppressed due to the tunneling between the
considered energy shell (\textrm{I}- and \textrm{C}-set) and other energy
levels. To further check the validity of the earlier analysis, we introduce
the average number fluctuation
\begin{equation}
\overline{\Delta n}_{\lambda ,j}=\lim_{T\rightarrow \infty }\frac{1}{T}%
\int_{0}^{T}\Delta n_{\lambda ,j}\mathrm{d}t,
\end{equation}%
to compare the results of the resonant and non-resonant cases. Average $%
\overline{r}$ and $\overline{\Delta n}_{\lambda ,j}$ as functions of
parameter $V$ for different $j$ are plotted in Fig. \ref{fig3}. It can be
shown that the dips appear in both $\overline{r}$ and $\overline{\Delta n}%
_{\lambda ,j}$ when the resonant electric field is taken. Another
interesting phenomenon is that both layers share the same conducting
behavior manifested by the performance of $\overline{\Delta n}_{a,j}$ and $%
\overline{\Delta n}_{b,j}$. This can be understood as follow: Assume that
the initial state is prepared as the direct product of ferromagnetic states
within each layer, i.e., $\left\vert \Uparrow \right\rangle _{\mathrm{A}%
}\left\vert \Downarrow \right\rangle _{\mathrm{B}}$. The resonant electric
field makes the spin-up particle migrates from the layer $\mathrm{A}$ to $%
\mathrm{B}$ such that a hole and a doublon are formed in two layers
individually. The spin-up particle can move freely based on the spin-down
ferromagnetic background while the formation of the hole enable the hopping
of the spin-down particles. Although the free particles are formed in
different ways in two layers, both facilitate the movement of particles
within each layer and thus increase the conductivity. Note that such a
steady state is the consequence of the non-equilibrium many-body dynamics
that is usually absent in the ground-state phase. Before ending this part,
it is worthy pointing out that the lattice structure of real materials, such
as bilayer graphene, is composed of coupled layers and is therefore not
strictly two-dimensional. Another advantage of our scheme is that it is not
only applicable to the bilayer system. Instead, it still makes sense even in
the multi-layer system. The core is that the interplay between the resonant
electric field and interaction can hybridize a small number of insulating
states and a large number of conducting states such that the particles tend
to move in each layer rather than localize at each site.

\section{Summary}

\label{summary}

In summary, we have systematically investigated the quench dynamics of the
bilayer Hubbard model subjected to a perpendicular electric field. In the
presence of the resonance electric field, the system can enter into the
steady conducting phase via a non-equilibrium scheme, which can be
demonstrated by the increase of the charge fluctuation. The underlying
mechanism is that the cooperation between the resonant electric field and
on-site interaction $U$ provides an energy shell hybridizing the single-and
double-occupied states with $s=0$. Due to the energy gap determined by the
large $U$ and resonant field, the inter- and intra- layer hopping does not
induce the tunneling between the considered shell and other shells so that
the dynamics in the quasi subspace is protected by the so-called "energy
conservation". The stable hybridization of such two kinds of states allow
the directional particle flow between the two layers, which forms the holes
in the \textrm{A} layer and doublon in the \textrm{B} layer. This
facilitates the movement of the particles within each layer and hence
enhances the conductivity. The particles exhibit the same conducting
behavior in both layers since the formation of the free particles shares a
similar mechanism in both layers. Due to the large proportion of
double-occupied states in the resonant energy shell, a non-equilibrium
system can accommodate such a conducting state for a long time.

Before ending this paper, we would like to explore two potential avenues for
future research based on the mechanism proposed in this study. The first
direction involves extending our analysis to bosonic systems. By considering
the presence of a resonant electric field, we anticipate that bosons will
also migrate from the upper layer to the lower layer, resulting in particle
fluctuations within each layer. This behavior can be interpreted as a potential signature indicating a transition from the Mott-insulating phase to a mixed phase characterized by conducting and superfluid behavior. The second future direction we propose is to introduce a harmonic driving field as a
replacement for the DC electric field employed in this study. Drawing
inspiration from the concept of Floquet engineering, the driving field would
provide photons that enable the insulating and conducting states to occupy a
common energy shell. This arrangement would facilitate intra- and
inter-layer transport within the system. Our findings present a promising strategy for enhancing the conductivity of multi-layer strongly correlated systems through an out-of-equilibrium
approach. These future research directions hold potential for further
advancing our understanding of the mechanisms governing conductivity and
phase transitions in complex systems.

\acknowledgments We acknowledge the support of the National Natural Science
Foundation of China (Grants No. 12275193, No. 11975166, and No. 11874225).


\end{document}